\documentclass[twocolumn,prb,aps,amsmath,amssymb,floatfix]{revtex4-1}
\usepackage{graphicx}
\usepackage{amsbsy}
\usepackage{color}
\DeclareGraphicsExtensions{.pdf}

\begin{document}
\title{Simulations of imaging of the local density of states by charged probe technique for resonant cavities}

\author{K. Kolasinski and B. Szafran}
 \affiliation{AGH University of Science and Technology, \\ Faculty of Physics and Applied Computer Science,
al. Mickiewicza 30, 30-059 Krak\'ow, Poland}

\date{\today}

\begin{abstract}
We simulate scanning probe imaging of the local density of states related to scattering Fermi level wave functions
inside a resonant cavity.
We calculate potential landscape within the cavity taking into account the Coulomb charge of the probe
and its screening by deformation of the two-dimensional electron gas using the local density approximation.
Approximation of the tip potential by a Lorentz function is discussed. The electron transfer problem
is solved with a finite difference approach.
We look for stable work points for the extraction of the local density of states  from conductance maps.
We find that conductance maps are highly correlated with the local density of states
when the Fermi energy level enters into Fano resonance with states localized within the cavity.
Generally outside resonances the correlation
between the local density of states and conductance maps is low.
\end{abstract}

\maketitle

\section{Introduction}

In semiconductor systems based on the two-dimensional electron gas (2DEG)  the linear conductance is determined by the scattering properties of the Fermi level wave functions.\cite{datta} Since relatively recently the electron transport properties can be probed with a local perturbation introduced to 2DEG by the charge probe at the atomic force microscope tip,\cite{sgmr1,sgmr2}
used as a floating gate.
The scanning gate microscopy \cite{sgmr1,sgmr2} allows for visualization of magnetic deflection of electron trajectories,\cite{g2} quantum  interference due to the elastic scattering \cite{g3} and Aharonov-Bohm effects,\cite{g4,g5} charged islands and the edge currents in nanostructurized quantum Hall bars,\cite{g7} branching of the electron flow,\cite{g8} tip-induced lifting of the Coulomb blockade in quantum dots,\cite{g9,g10} etc.

In this paper we consider electron transport across a cavity side-attached to the semiconducting channel.
The electron flow through resonant cavities is a basic problem for the quantum transport with a long history,
starting from the weak localization effects,\cite{bar} the relation between the classical and quantum modes of transport,\cite{akis} in particular the scars of the
classical trajectories on  wave functions, \cite{sc1,sc2} and most recently
 the pointer states \cite{zurek,qda1} that are robust against decoherence and stable despite coupling to the environment.
The spatial distribution of the pointer states was extracted by post-treatment of the conductance images as obtained by scanning gate microscopy.\cite{sgmr1,g6}

In  this work we consider a purely coherent electron transport. Our purpose is to determine an extent to which the details of the local density of states \cite{g4,g5} at the Fermi level
can be extracted from the raw conductance maps gathered with the charge perturbation scanning the surface of the structure.
The original tip potential as seen by the 2DEG is of the long range Coulomb form. The Coulomb potential is screened by deformation of 2DEG density.
Usually for theoretical modeling the tip potential is assumed short-range in a form given by a closed formula.\cite{g2,g4,g5,g12,g14}
For 2DEG which is not confined laterally \cite{g2,g3,g8} the deformation of the electron gas follows the tip as it scans the surface and the effective (screened) tip potential preserves its form. On the other hand in systems with lateral confinement -- in the cavity in particular -- the 2DEG deformation cannot freely follow the tip, so a local form of the effective potential can only be an approximation
of the actual potential.
In this work the effective potential of the tip is evaluated by solving the density functional theory equations.\cite{g11}
 We find that the tip potential is close to Lorentzian which for small 2DEG-tip distance is isotropic outside the edges of the cavity
and of the width which is close to the tip-2DEG distance.

We demonstrate that the resolution of the local density of states at the Fermi with the charge probe technique
depends on the work point defined by the electron density. The local density of states is highly correlated
to conductance maps when the Fermi energy level enters into Fano resonance with states localized within the cavity.
Outside the resonances the maps calculated for weak tip potentials agree with the Lipmann-Schwinger perturbation theory \cite{lpst}
but are not correlated in a clear way with the local density of states.

\section{Theory}

\begin{figure}[ht!]
\hbox{
	   \includegraphics[width=80mm]{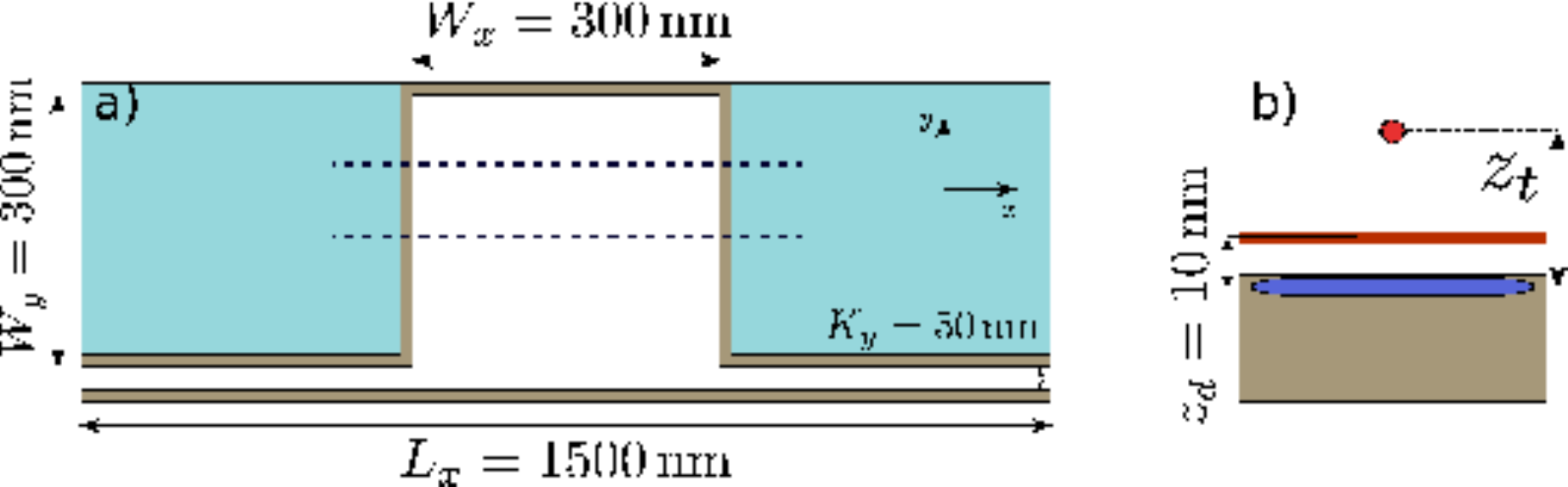}
          }
\caption{ Schematics of the considered system. (a) Geometry within the plane of confinement.
The square cavity of side length 300 nm is side attached to 50 nm wide channel.
The length of the computational box is 1.5 $\mu$m. The dashed straight lines indicate the paths
along which the Lorentzian fit to the tip potential is discussed (see Fig. 3 below).
 (b) Side view: the plane of
electron confinement at a distance of $z_d$ to the sheet of ionized donors and $z_t$ to the tip.}
\label{SCHEMAT}
\end{figure}

We consider the electron gas filling the channels including  the side-attached square cavity ($300$ nm $\times 300$ nm),
that is depicted in Fig. \ref{SCHEMAT}(a). The entire computational box is taken as large as 1.5$\mu$m.  This length guarantees,
that the potential of the tip is screened before it reaches the ends of the channels. The channels are taken 50 nm wide. The electron gas
is considered strictly two-dimensional. We apply local density approximation (LDA)  for description of the electrostatics of the system, \cite{stopa}
with the single-electron Hamiltonian
\begin{equation}
H=-\frac{\hbar^2 \nabla^2}{2m_\mathrm{eff}}+U(x,y)\label{Uxy}
\end{equation}
where $m_\mathrm{eff}$ is the electron effective band mass,
and the potential is given by,
\begin{equation}
U=W+V_H+V_{xc} +V_{tip},
\end{equation}
where $W$ is the confinement potential of the channels (we take 0 inside the channels and 200 meV outside),
$V_H$ is the Hartree potential, $V_{xc}$ -- the exchange correlation potential (we apply
the parametrization by Perdew and Zunger\cite{perdz}), and $V_{tip}$ is the tip potential.
The Coulomb potential of the charge at the tip of the probe as seen at the 2DEG level is given by
\begin{equation}
V_{tip}(x,y;x_t,y_t)=\frac{eQ_{tip}}{4\pi\epsilon\epsilon_0\sqrt{(x-x_t)^2+(y-y_t)^2+z_t^2}},\label{ac}
\end{equation}
where $(x_t,y_t,z_t)$ is the tip location, and $Q_{tip}$ is the charge at the tip.

In 2DEG, the carriers are delivered by the donor layer which is left charged positively.
We assume that a homogenous sheet of positive charge is present on top of the area accessible to electrons [see Fig. \ref{SCHEMAT}(b)] at a distance of $z_d$
from the electron gas.
The considered model is  charge neutral,
 with the
average electron density within the channels matched to the donor density
in the doped layer above the 2DEG [see Fig. \ref{SCHEMAT}(b)].

In order to simulate open infinite leads we apply periodic boundary conditions for the Schroedinger equation
with Hamiltonian (\ref{Uxy}). For calculation of the Hartree potential we produce copies of the charge present
within the original computational box and place them on its left and right sides [see Fig. \ref{SCHEMAT}(a)], and then integrate
the charge with the Coulomb potential
\begin{eqnarray}
    V_H(x,y) &=& \frac{e^2}{4\pi\epsilon\epsilon_0} \left(-\int dx'dy' \frac{n_d(x',y',z_d)}{|(x,y,0)-(x',y',z_d)|}  \right. \nonumber \\&&
    \left. +\int dx'dy'  \frac{n(x',y',0)}{|(x,y,0)-(x',y',0)|} \right) \label{V_H}
    \end{eqnarray}
In Eq. (\ref{V_H}), $n$ is the electron density, which is determined in the following manner.
We solve the eigenequation for $H$ obtaining eigenfunctions $\psi_{i}$
with eigenvalues $\varepsilon_{i}$. Then, the electron density is calculated as \begin{equation}
n(\boldsymbol{r})=2\sum_{i}^{\infty}f(\varepsilon_{i})|\psi_{i}(\boldsymbol{r})|^{2},
\end{equation}
where
$f$ is given by the Fermi-Dirac distribution,
\begin{equation}
f(\varepsilon_{i})=\left[\exp\left(\left({\varepsilon_{i}-E_{F}}\right)/{k_{B}\tau}\right)+1\right]^{-1},
\end{equation}
and the factor 2 accounts
for the degeneracy of energy levels with respect to the spin.
The Fermi energy $E_F$ is determined by the normalization condition
\begin{equation}
N=2\sum_{l}^{\infty}f(\varepsilon_{l}).\label{noninteger}
\end{equation}
In the calculations we assume the temperature of $\tau=0.28$ K and material parameters of InAs: $m_\mathrm{eff}=0.023 m_0$ and  $\epsilon=15.5$.

Once the self-consistence of the Schroedinger-Poisson scheme is reached, we obtain the Fermi energy and the potential distribution which is then
used for determination of the electron transfer probability $T$, and thus the conductance $G=\frac{2e^2}{h}T$,
according to the Landauer approach.
Within the leads, far away from the cavity the potential depends only on $y$, so that
the Hamiltonian eigenstates carrying the probability current can be described
by the wave vector ($k>0$),
\begin{equation}
\Psi_k(x,y)=e^{ikx}\psi(y),\label{spin}
\end{equation}
where $\psi$ is the wave function describing the state of the transverse quantization.
In this paper we consider low electron densities, for which the transport occurs in the lowest subband only.
Then a single value of the wave vector $k$ appears
at the Fermi level.
In order to determine the electron transfer probability we solve the Schroedinger equation with the finite difference
method and the scattering boundary conditions. Let us assume that the electron
is incident from regions of negative $x$ (see Fig. 1) and let us
denote the scattering wave function for this case by $\Psi_+$.
In the channels far away
from the cavity -- i.e. outside the range of the evanescent modes -- the wave function
acquires the form
\begin{equation}
\Psi_+(x,y)=\Psi_{k}(x,y)+r_+ \Psi_{-k}(x,y)\label{aa}
\end{equation}
in the input channel, where $r_+$ is the backscattering amplitude.
In the output channel one finds
\begin{equation}
\Psi_+(x,y)=t_+\Psi_{k}(x,y)\label{aaa},
\end{equation}
where $t_+$ is the electron transfer amplitude for the electron incident from the left. The electron transfer probability is $T=|t_+|^2$.
The solution of the scattering problem, i.e. the Hamiltonian eigenstate matched to boundary conditions given by Eqs. (\ref{aa}) and (\ref{aaa}) is found by an iterative approach which is described in detail in Refs. \onlinecite{szp,g11}.

Below we discuss the spatial density of states at the Fermi level (also known as the local density of states -- LDOS \cite{g4}).
The local density of states is obtained as a sum of the scattering probability
densities for the Fermi-level-electron incident to the cavity from the left $\Psi_{+}$ and right channels $\Psi_{-}$
\begin{equation}
\mathrm{LDOS}(x,y)=|\Psi_{+}(x,y)|^2+|\Psi_{-}(x,y)|^2.
\end{equation}

In order to support the discussion of the simulated scanning gate microscopy conductance maps, in the limit of weak perturbations
we employ the
general formulas for the first and second order corrections to conductance due to the tip  developed \cite{lpst} using the Lippmann-Schwinger
approach.
In the lowest-subband transport case discussed here,
the formulas read
\begin{equation}
G^{(1)}(x_t,y_t)=-\frac{2m_\mathrm{eff}}{\hbar^2 k}\Im(r_+^*t_-V^{(-,+)}), \label{fop}
\end{equation}
for the first-order
and
\begin{eqnarray}
G^{(2)}(x_t,y_t)&=&-\frac{2\pi m_\mathrm{eff}}{\hbar^2 k} \Re\left[|t_+V^{(-,+)}|^2   +|t_-V^{(+,-)}|^2  \right. \\ \nonumber&&\left. +r_+^*t_-(V^{(-,-)}V^{(-,+)}-V^{(-,+)}V^{(+,+)}) \right] ,\label{sop}
\end{eqnarray}
for the second-order corrections,
with the potential matrix element defined as
\begin{equation} V^{a,b}(x_t,y_t)=\int dx dy \Psi_a^*(x,y)V(x,y;x_t,y_t) \Psi_b(x,y) \label{eme} \end{equation}
that is evaluated from the scattering wave functions coming from left ($a,b=+$) or right ($a,b=-$) leads.
The wave functions and the scattering amplitudes used in Eqs. (13-15) are calculated in the absence of the tip,
which only enters the kernel of Eq. (\ref{eme}).
The conductance is then approximated by $G_{pert}(x_t,y_t)=G(\infty)+G^{(1)}(x_t,y_t)+G^{(2)}(x_t,y_t)$,
where $G(\infty)$ is the conductance in the absence of the tip.

For quantitative discussion of the similarity between maps of conductance
and the local density of states we calculate
correlation factors in the following manner.
We first perform normalization of the map
\begin{equation}
f=\frac{F-F_{min}}{F_{max}-F_{min}},
\end{equation}
where $F_{max}$ and $F_{min}$ are the maximal and minimal values
of the quantity $F$ within the cavity area $S$.
Next, we calculate
the correlation factor for dimensionless normalized maps $f$ and $g$ as
\begin{equation}
r=\frac{1}{S}\frac{\int_S dxdy (f(x,y)-\langle f \rangle)(g(x,y)-\langle g \rangle) }{\sigma_f \sigma_g}, \label{corfac}
\end{equation}
where
 $\langle f\rangle=\int_S dxdy f(x,y)/S$ and  $\sigma_f^2 =\int_S dxdy (f(x,y)-\langle f\rangle )^2/S$.

\begin{figure}[ht!]
\hbox{
\begin{tabular}{c}
           	   \includegraphics [width=80mm]{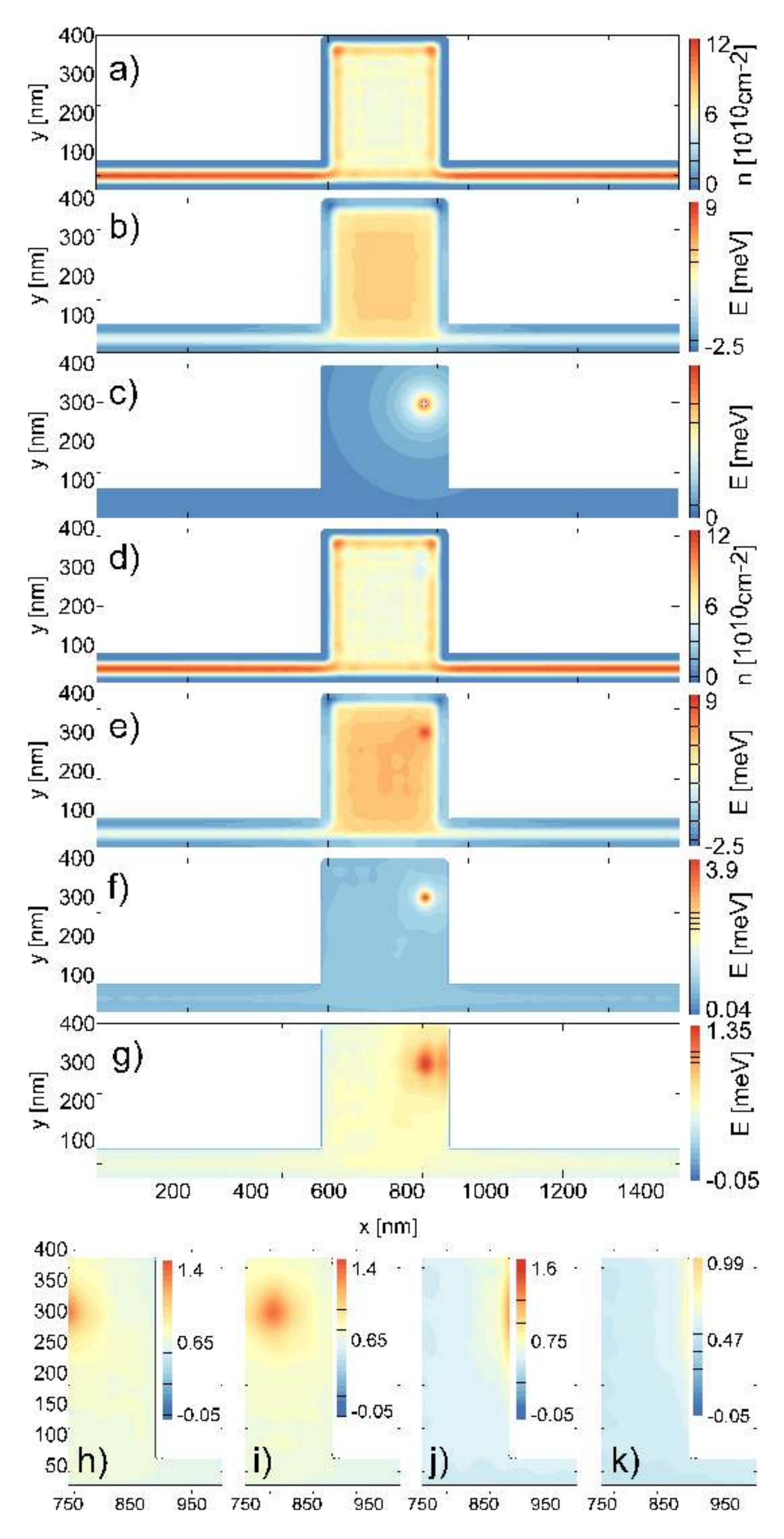}
  \end{tabular}
          }
\caption{
(a) Charge density of the electron gas for 100 electrons present within
the computational box in the absence of the tip. (b) Total potential $U$ distribution
corresponding to (a). (c) Bare potential of the tip
within 2DEG plane. The probe charge is above the spot marked by the cross for $Q_{tip}=-1$ [e] at a distance of $z_t=15$ nm
from the plane. (d) Charge density for the tip localized like in (c).
(e) The potential $U$ distribution
corresponding to (d). (f) The effective potential of the tip calculated as the difference of (e) and (b).
(g) The effective potential for $z_t=40$ nm for $y_t=300$ nm and $x_t=850$ nm.
(h,i,j,k) same as (g) only for $x_t=750$ nm, $800$ nm, 950 nm,  and 1000 nm, respectively.
}

\label{pe}
\end{figure}

\begin{figure}[ht!]
\hbox{
\begin{tabular}{c}
           	   \includegraphics[width=70mm]{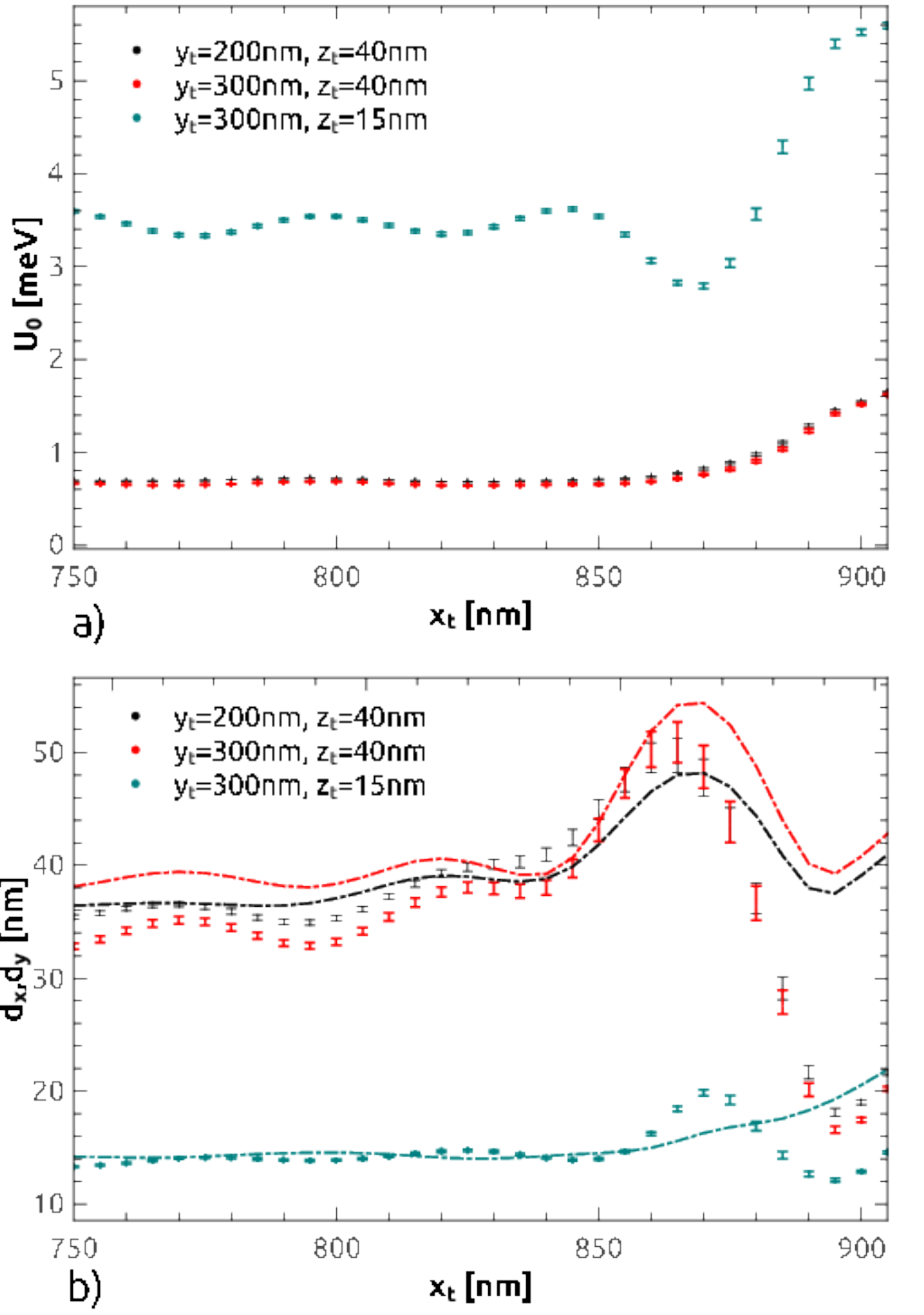}
  \end{tabular}
          }
\caption{
Parameters of the Lorentz potential [Eq. \ref{mo}] fitted to the effective
tip potential for $Q_{tip}=-|e|$ and the tip-2DEG distance of $z_t=15$ nm (as in Fig. 2(e)] and $z_t=40$ nm
along the dashed lines plotted in Fig. 1. Plot (a) shows the height of the potential $U_0$
maximum. Plot (b) shows $d_y$ (dash-dotted curves) and $d_x$ given by the error bars.
The fit was performed within the cavity area inside a square of 200 nm side length centered on the tip.
}
\label{ls}
\end{figure}

\section{Results and Discussion}

\subsection{Screening of the tip potential}
The calculated self-consistent charge and potential distributions as obtained for $N=100$ electrons
within the box in the absence of the tip
are plotted in Figures \ref{pe}(a) and \ref{pe}(b), respectively.
The charge density within the system is distinctly maximal at its edges.
In the channels near the ends
of the computational box $U$ becomes independent of $x$, which is consistent with the boundary conditions  used for
the transport problem [Eq. (\ref{spin})].
Figure \ref{pe}(c) shows the bare potential of the tip localized at $z_t=15$ nm above the 2DEG.  The applied effective charge at the tip $Q_{tip}=-1$ [$e$] for the
tip radius of $R=5$ nm corresponds to the tip potential of $V=Q_{tip}/R=0.288$ V.
Figures \ref{pe}(d) and \ref{pe}(e) display the charge density and potential as perturbed by the tip, respectively.
The consequence of the charge redistribution is the screening of the original tip potential.
The screened (effective) tip potential can be calculated as a difference of perturbed [Fig. \ref{pe}(e)] and unperturbed potentials [Fig. \ref{pe}(b)].
The difference is displayed in Fig. \ref{pe}(f). The effective potential has a form of a 4 meV peak
of $\simeq$ 30 nm diameter. For comparison, the effective potential calculated for the 2DEG-tip distance
increased to 40 nm is displayed in Fig. \ref{pe}(g-l).
As the tip approaches the edge of the cavity the second elongated maximum is formed along the edge [Fig. \ref{pe}(g)].
In the present model the screening is only due to 2DEG which is missing outside the cavity hence
the reduced screening at the edge.
When the tip is displaced to the outside of the cavity [Fig. \ref{pe}(j,k)]
the elongated maximum at the edge is preserved.

For a quantitative parametrization of the tip potential we fitted
to the results of the Schroedinger-Poisson scheme the Lorentz function
\begin{equation}
V_L(x,y;x_t,y_t)=\frac{U_0}{1+\frac{(x-x_t)^2}{d_x^2}+\frac{(y-y_t)^2}{d_y^2}}\label{mo},
\end{equation}
where $U_0$ is the potential height above the constant potential background, $d_x$ and $d_y$ are potential widths in the directions parallel
and perpendicular to the channels, respectively.
The results of the fit
for the tip following the paths marked by horizontal dashed lines in Fig. 1
are displayed in Fig. \ref{ls} for parameters of Fig. 2.
For stronger interaction ($z_t=15$ nm) the height of the tip potential oscillates. The period of the oscillation
is roughly half of the Fermi wavelength (here $k_F=0.05$/nm), which is reminiscent of the
Friedel oscillations discussed in Ref. \onlinecite{g11}.
When the tip is near the center of the cavity [cf. Fig. 2(b) for $y=300$ nm
and $x_t=750$ nm] its potential is isotropic. Note that, in this case the potential width $d_x=d_y$ is of the order of $z_t$.
The height of the tip potential [Fig. \ref{ls}(a)] tends to increase at the edge of the cavity which
is due to the reduced screening by the electron gas.
The potential becomes anisotropic when the tip approaches the edges of the cavity due
to the anisotropy of the screening medium.
An extreme anisotropy is reached for the tip
above the edge: the potential is strongly elongated along the edge, it weakly penetrates
the cavity [cf. Fig. 2(j)].

For the discussion below we use both the results for the Coulomb potential (\ref{ac}) and the anisotropic Lorentz
ansatz (\ref{mo}) with $d=d_x=d_y$. There is an essential difference between these two calculations.
$V_{tip}$ given by Eq. (\ref{ac}) enters the DFT equations,
so that the charge density $n$ reacts to its presence resulting in the screening
of the tip potential.
On the other hand, whenever the ansatz in form given by  Eq. (\ref{mo}) is used,
we simply add this potential to the self-consistent potential as obtained
by DFT in the absence of the tip and use it in the electron scattering problem.

The present finding that the tip potential can be approximated by the Lorentz function
with the width of the order of the 2DEG - tip distance is consistent with the experiment of Ref. \onlinecite{g10}
which investigated the form of the actual tip potential
in the Coulomb blockade microscopy of double quantum dots defined within 2DEG. The conclusion \cite{g10} was
that the potential is close to the circularly symmetric Lorentz function that for the tip $\simeq 234$ nm above the electron gas (200 nm above
the sample surface with the electron gas buried 34 nm below it)
possesses a dip that is 250 nm wide.

\begin{figure}[ht!]
\hbox{
         	   \includegraphics [width=90mm]{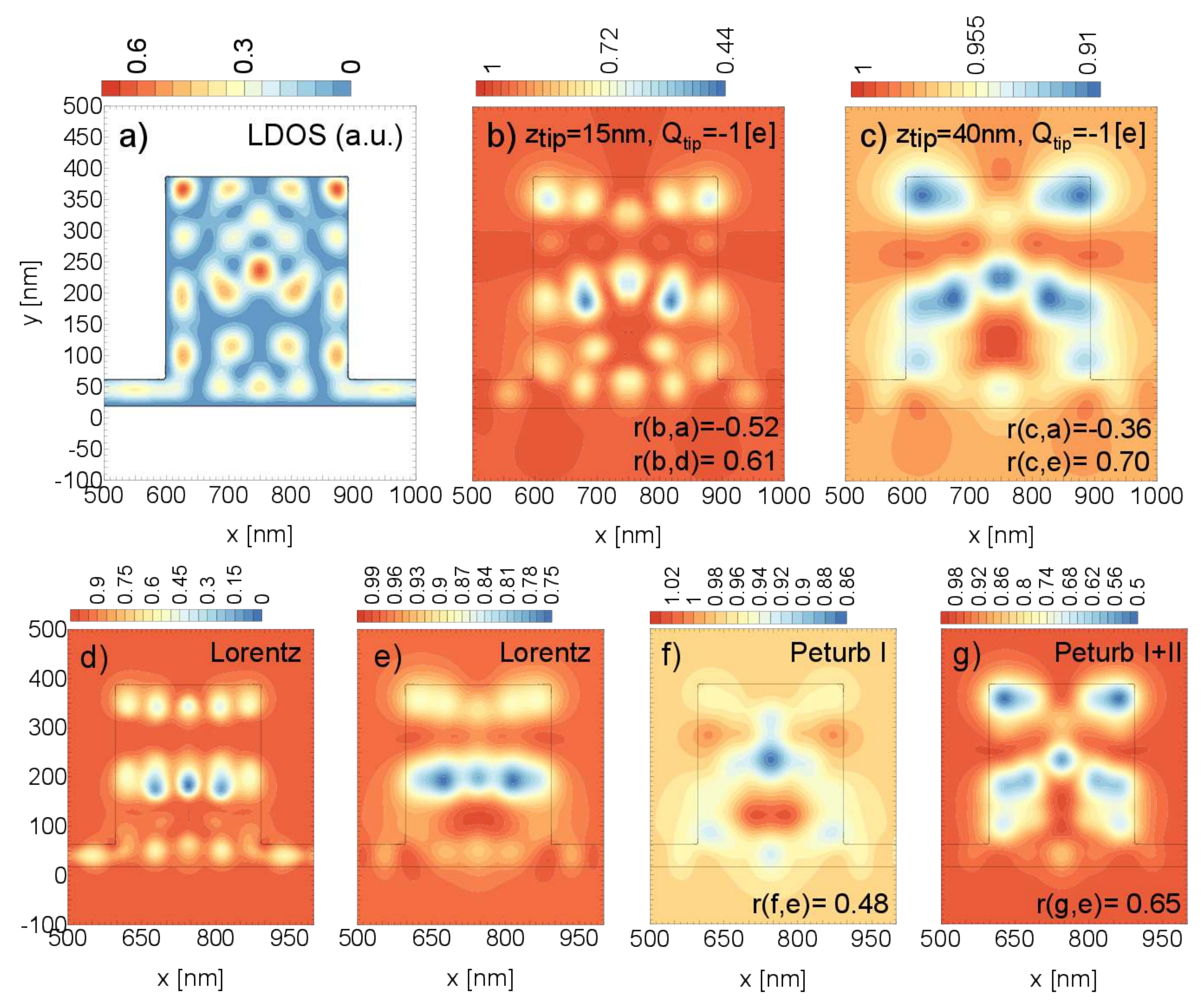}
          }
\caption{Results for $N=78$ electrons within the computational box.
(a) The density of states at the Fermi level (local density of states), calculated as a sum of probability densities
corresponding to scattering wave functions incident with $k_F=0.039$ / nm ($\hbar ^2 k_F^2/{2m_\mathrm{eff}}=2.5$ meV) from the left and right
to the cavity. Transfer probability maps [or conductance maps in units of $G_0=2e^2/h$]
obtained for $Q_{tip}=-1 [e]$ and the tip at the distance of $z_t=15$ nm (b) and $z_t=40$ nm (c) from the 2DEG.
(d) Conduction map for Lorentz perturbation [Eq. \ref{mo}] with parameters corresponding to plot (b), i.e.
$U_0=3.5$ meV and $d=d_x=d_y=15$ nm. (e) Conduction map for Lorentz perturbation with parameters corresponding to plot (c), i.e.
$U_0=0.7$ meV and $d=40$ nm. Plots (f) and (g) show the conductance
maps as calculated with the first and the second order corrections of the Lipmann-Schwinger equation theory [Eqs. (\ref{fop}, \ref{sop})]
obtained with the potential of plot (e).
The values of $r(b,a),r(b,d),\dots $ etc. give the correlation factors between
the maps of panels $(b,a),(b,d),\dots$, as calculated according to Eq. (\ref{corfac}).
}
\label{zt}
\end{figure}

\begin{figure}[ht!]
\hbox{
         	 \includegraphics[width=70mm]{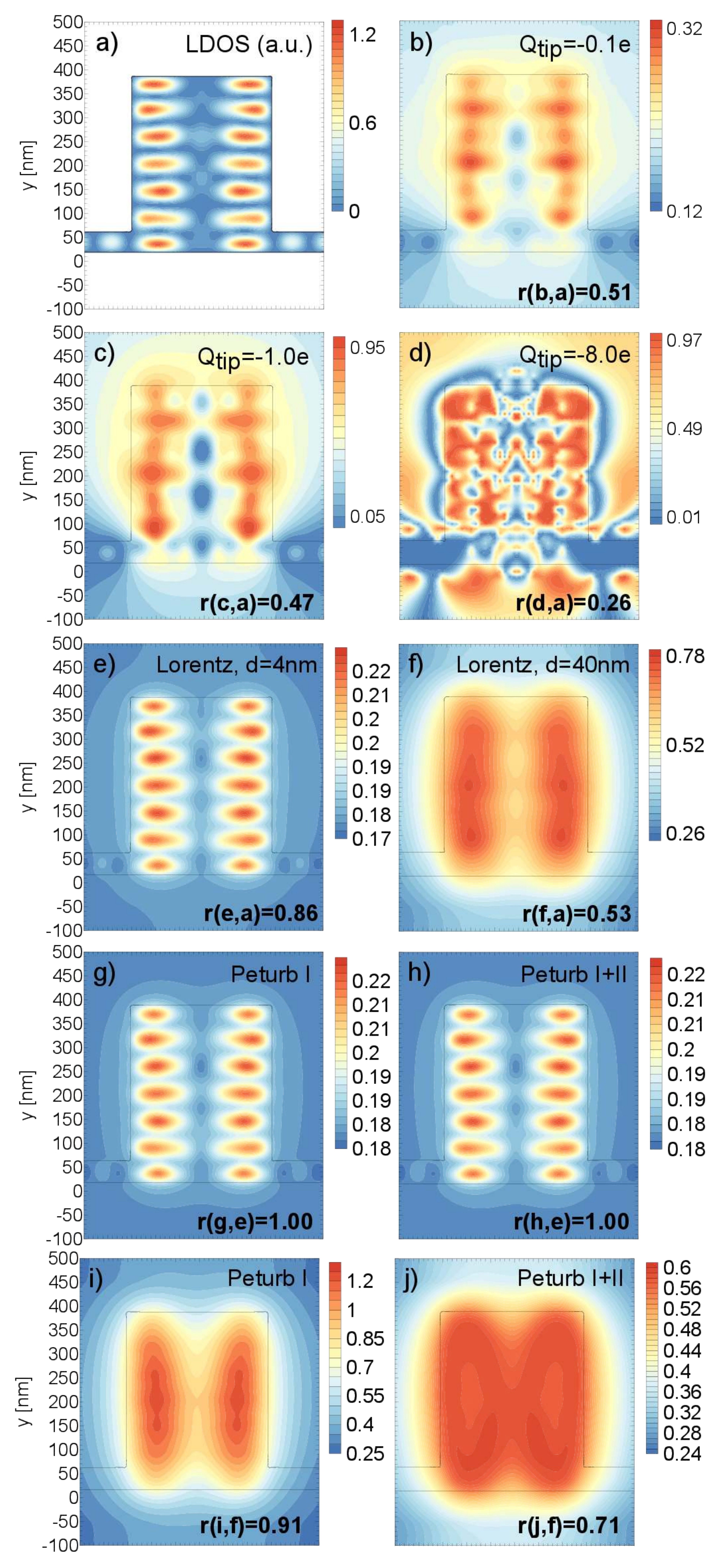}
          }
\caption{Results for $N=92$ electrons within the computational box ($k_F=0.046$ /nm, $\hbar ^2 k_F^2/{2m_\mathrm{eff}}=3.5$ meV)
 (a) The density of states at the Fermi level (local density of states), and conduction maps obtained for the tip at the distance of $z_t=15$ nm,
for  $Q_{tip}=-0.1$ [e] (b), $Q_{tip}=-1 [e]$ (c), and $Q_{tip}=-8$ [e] (d).
Results for a Lorentzian ansatz of the effective tip potential are displayed in (e) and (f)
for $U_0=0.33$ meV and radii $d=4$ nm (e) and $d=40$ nm (f).
Plots (g) and (h) [(i) and (j)]  are the first and second-order perturbations
calculated for parameters of (e) [(f)].
}
\label{q92}
\end{figure}

\subsection{Density of states at the Fermi level versus the conduction maps}

In Fig. \ref{zt}(a) we show the density of states at the Fermi level obtained for $N=78$
electrons within the computational box. Figures \ref{zt}(b) and \ref{zt}(c) display the
corresponding maps of the transfer probability for $z_t=15$ nm and $z_t=40$ nm, respectively.
Correlation between the LDOS and $T$ map for $z_t=15$ nm is evident, although the relative
amplitudes of extrema as observed in LDOS and $T$  vary. The correlation factor between
the $T$ map with $z_t=15$ nm and the LDOS [Figs. \ref{zt}(a) and (b), respectively]
is $r=-0.52$ (the negative sign is due to the coincidence between maxima of LDOS and minima of $G$ map).
The details of the $T$ map become less resolved for $z_t=40$ nm [Fig. \ref{zt}(c)] as should be expected from the increase of the effective width of the tip discussed in Fig. \ref{pe}(f) and Fig. \ref{pe}(g).
The correlation factor between the $T$ map of Fig. \ref{zt}(c) and the LDOS of Fig. \ref{zt}(a) falls to $r=-0.36$.

As discussed in Fig. \ref{ls}, the parameters of the effective tip potential vary with the tip position.
In order to illustrate the importance of the variation in Fig. \ref{zt}(d) and (e) we presented
the $T$ maps as obtained with the isotropic Lorentz potential [Eq. \ref{mo}].
For $z_t=15$ nm ($z_t=40$ nm) we fixed parameters  to
$U_0=3.5$ meV and $d=d_x=d_y=15$ nm ($U_0=0.7$ meV and $d=40$ nm)
according to the fit obtained for the tip near the center of the cavity [cf. Fig. \ref{ls}].
The similarity between the $T$ maps
with the Lorentz ansatz [Fig. \ref{zt}(d,e)] and the ones obtained by the full calculation with the Coulomb potential
of the tip [Fig. \ref{zt}(b,c)] is quite distinct and the correlation factor equals $r=0.61$ for $z_t=15$ nm and $r=0.7$ for $z_t=40$ nm.
For $z_t=40$ nm, the results with the Coulomb potential indicate a stronger variation along
the edges of the cavity which results from the increased height of the actual perturbation near the edges
(see the discussion above).

For the parameters corresponding to $z_t=40$ nm the height of the perturbation $U_0=0.7$ meV is distinctly
smaller than the kinetic energy of the Fermi level electron ($\hbar ^2 k_F^2/{2m_\mathrm{eff}}=2.5$ meV).
It is instructive to check how the perturbation theory works for the Lorentz potential
as compared to the exact results [Fig. \ref{zt}(e)]. The first and second order
$T$ maps are displayed in Figs. \ref{zt}(f) and (g), with the correlation factors
reaching $r$=0.48 and 0.65, respectively.

The correspondence of LDOS with the $T$ map found in Fig. \ref{zt} for $z_t=15$ nm  is relatively close.
In order to see whether this can be considered a rule, we performed calculations for a varied electron
number -- $N=92$ -- see the results of Fig. \ref{q92}.
The result for $Q_{tip}=-1[e]$ [Fig. \ref{q92}(c)] exhibits
 a similar correlation to LDOS [Fig. \ref{q92}(a)] with the correlation factor of $r=0.47$.
Note, that the correspondence has now an {inverse} character with respect to Fig. \ref{zt}:
in Fig. \ref{q92} the maximal $T$ corresponds to maximal LDOS hence the positive sign of $r$.
Figure \ref{q92}(d) indicates that the correspondence of $T$ to LDOS is reduced for the larger charge at the tip. For
$Q_{tip}=-8e$ the 2DEG beneath is completely depleted, and the volume of the cavity itself, i.e. the
space accessible for electrons changes with the varied tip position, hence the abrupt variation of the $T$ map.
The correlation to LDOS factor drops to $r=0.26$.

We learn from Fig. \ref{q92}(b) that as the tip charge is reduced further
from $Q_{tip}=-e$ to $Q_{tip}=-0.1e$  the $T$ map does not get any closer to LDOS. Only the amplitude of the variation is reduced
for smaller $|Q_{tip}|$.
We find that in general, the effective width of the tip depends on $z_t$ and not on $Q_{tip}$.
The experimental literature on quantum rings \cite{g4,sgmr1} contains the discussion of the influence
of the tip potential on the conductance patterns in the scanning gate microscopy maps.
The conclusion \cite{sgmr1,g4} was that in the linear perturbation regime (i.e. for relatively weak perturbation) the potential at the tip
influences only the amplitude of the map and not the pattern, which is consistent with the result
obtained for $Q_{tip}$ varied between $-e$ and $-0.1e$.

For comparison in Figs. \ref{q92}(e) and (f) we presented results for $T$ maps as obtained for the Lorentz potential of the tip
for $d=4$ nm and $d=40$ nm, respectively for a small height of potential $U_0=0.33$ meV (about 10\% of $\frac{\hbar^2k_F^2}{2m_\mathrm{eff}}$).
The result of Fig. \ref{q92}(e) for $d=4$ nm reproduces in a close detail the LDOS of Fig. \ref{q92}(a) -- with
the correlation factor reaching $r=0.86$. The result with $d=4$ nm and small $U_0$ is exactly
reproduced by the perturbation calculus [Fig. \ref{q92}(g-h)]. For the wider tip $d=40$ nm,
the second correction plays a more decisive role [Fig. \ref{q92}(i,j)],
with the values of $T$ closer to the results of the exact calculation [Fig.\ref{q92}(f)] but
with a reduced correlation of the images due to lower contrast of the map Fig.\ref{q92}(j).
The discussed case of $d=40$ nm is near the verge of the applicability of the perturbation theory.
We find that independent of $N$ for $d=4$ nm and $U_0$ of the order of 10\% of the kinetic Fermi energy the
perturbation calculus \cite{lpst} including the second order correction exactly reproduces the exact transfer probability maps with $r$ close to 1.

\begin{figure}[ht!]
\hbox{
\begin{tabular}{l}
         	   (a) \\ \includegraphics[width=75mm]{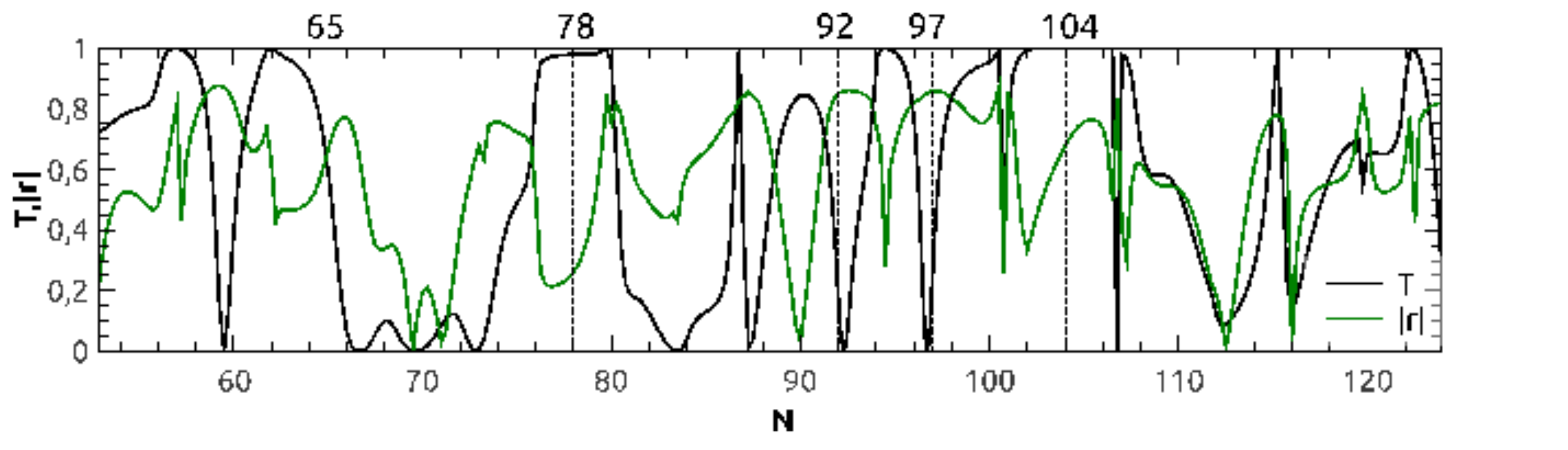} \\
         \end{tabular}
                   }
\caption{
The black line shows the dependence of the transfer probability $T$ on $N$.
The green curve shows the correlation factor between LDOS and the $T$ map
as obtained for a point-like perturbation $d=4$ nm and $U_0=0.1\frac{\hbar^2 k_F^2}{2m_\mathrm{eff}}$.
The vertical dashed lines show the electron numbers which are discussed in the text.}
\label{todn}
\end{figure}

\begin{figure}[ht!]
\hbox{
\begin{tabular}{c}
         	   \includegraphics[width=80mm]{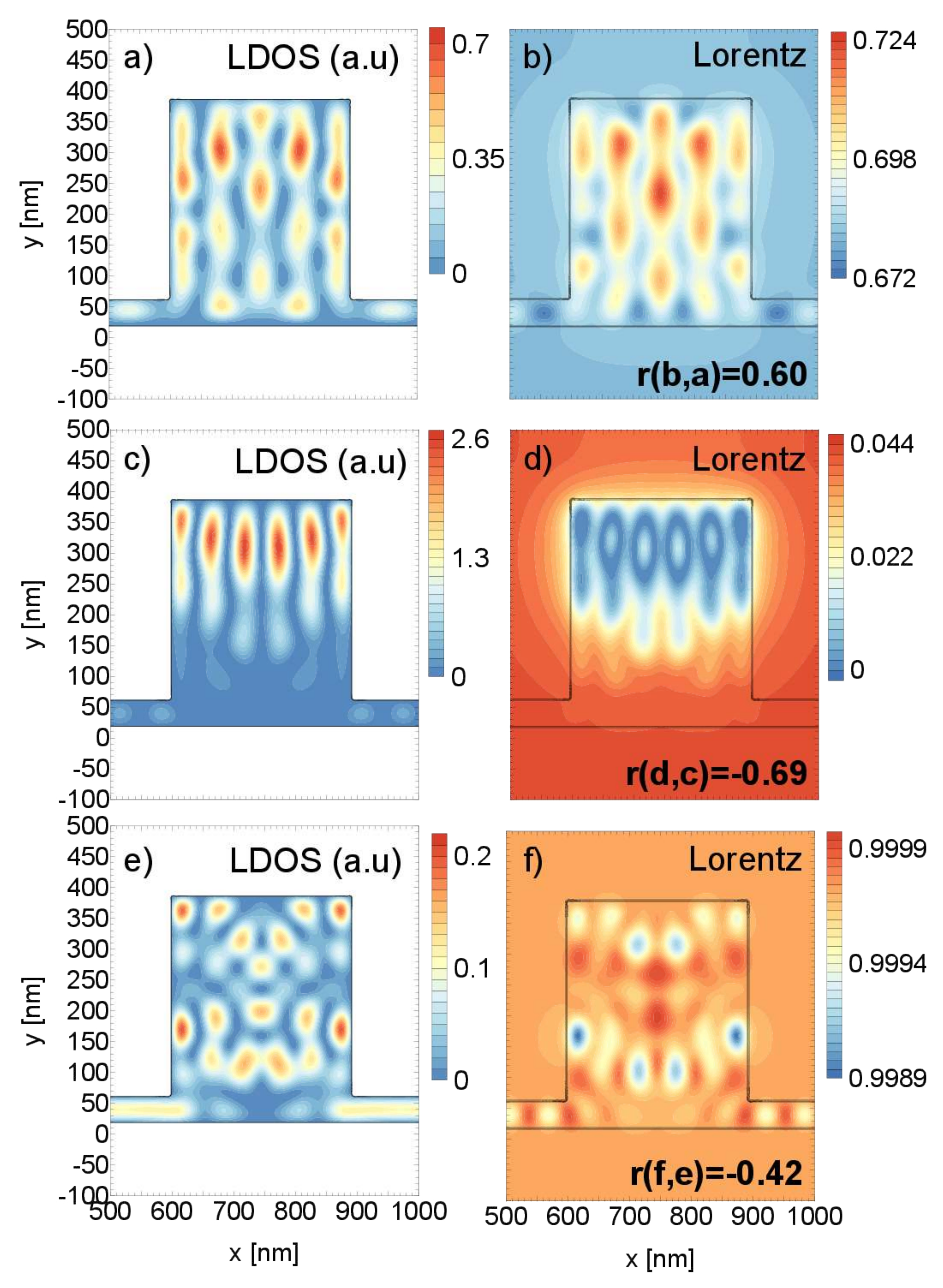}

         \end{tabular}
          }
\caption{
The first row of plots corresponds to $N=65$, second to $N=97$, and the third to $N=104$ within the system.
In the first column we show the local density of states,
and in the second column the $T$ maps as obtained with the Lorentz ansatz of the effective tip potential for $U_0=0.1 \hbar^2/k_F^2/2m\mathrm{eff}$ and width $d=4$ nm.
The Fermi wave vectors and the related kinetic energies for $N=65$, 97 and 104 are: 0.032 / nm, 0.049 / nm, 0.052 /nm; 1.7 meV, 4 and 4.5 meV, respectively.
 }
\label{ds}
\end{figure}

\subsection{$T(N)$ versus the resolution of the local density of states at the Fermi level}

LDOS -- that characterizes only the scatterer in the absence of the tip -- can be resolved by scanning gate microscopy maps with the spatial precision that is limited by the width of the tip potential.
When the tip potential is large, the scattering wave functions will be strongly influenced by the tip.
In the limit of weak and point-like tip perturbation the correlation between the LDOS and the $T$ maps should be as high as possible.
For the rest of the paper we concentrate on this limit. The limit of weak and delta-like perturbation
is investigated using a short range Lorentz potential $d=4$ nm and small height $U_0=0.1 {\hbar^2k_F^2}/{2m_\mathrm{eff}}$.

We find that the resolution of the local density of states near the Fermi level by the $T$ maps depends on the electron density (or $N$) which sets the Fermi energy within the system.
The dependence of $T$ on $N$ is displayed in Fig. \ref{todn}.
For the purpose of Fig. \ref{todn},  we allowed $N$ to take on also non-integer values
[$N$ enters the expression for the Fermi energy (\ref{noninteger})].
Additionally, we plotted the correlation factor between LDOS and the $T$ map as calculated
with the Lorentz potential.

The results for $N=78$ presented above [Fig. \ref{zt}(b)]
corresponded to a plateau of $T(N)$ in Fig. \ref{todn}. In this case
the correlation factor $r$ between the point-like weak perturbation is as small as 0.35.
The case of $N=92$  in Fig. \ref{q92} corresponds to a neighborhood of a dip of $T$ [see Fig. \ref{todn}(a,c)].
This dip results from the Fano interference that involves a resonant localized state within the cavity.
The correlation factor becomes as large as $r=0.86$.
When $|dT/dN|$ is large, so is the amplitude of the $T(x,y)$ variation as obtained for the Coulomb tip with $Q_{tip}=-e$ at $z_t=15$ nm.
For maps plotted at the plateaux of $T$ the amplitude
is distinctly smaller [see: Fig. \ref{zt}(b)].

We find as a general rule, that near the Fano resonances the correlation between
the LDOS and $T$ maps reaches $r\simeq 0.8$ (see $N=59$, 92.3, 96.7, 106.7). For an extra illustration the results near another resonance for $N=97$ are displayed in Fig. \ref{ds}(c,d).
Near the plateaux of $T$ the correlation becomes distinctly lower -- see
the results for $N=78$ discussed above or
the results presented for $N=104$  in Fig. \ref{ds}(e,f).
Note, that for $N=104$ there is an apparent similarity between the LDOS and the $T$ map
since the extrema of the maps coincide with one another. However,
the maxima of LDOS correspond variably to either minimum or maximum of $T$ -- hence
the low value of $r=-0.42$.
The case for $N=65$ of Fig. \ref{ds}(a,b) corresponds to an off-resonant
conditions but for large $|dT/dN|$. In this case the correlation between the LDOS and the conductance map is still significant.

\begin{figure}[ht!]
\hbox{
         	   \includegraphics[width=90mm]{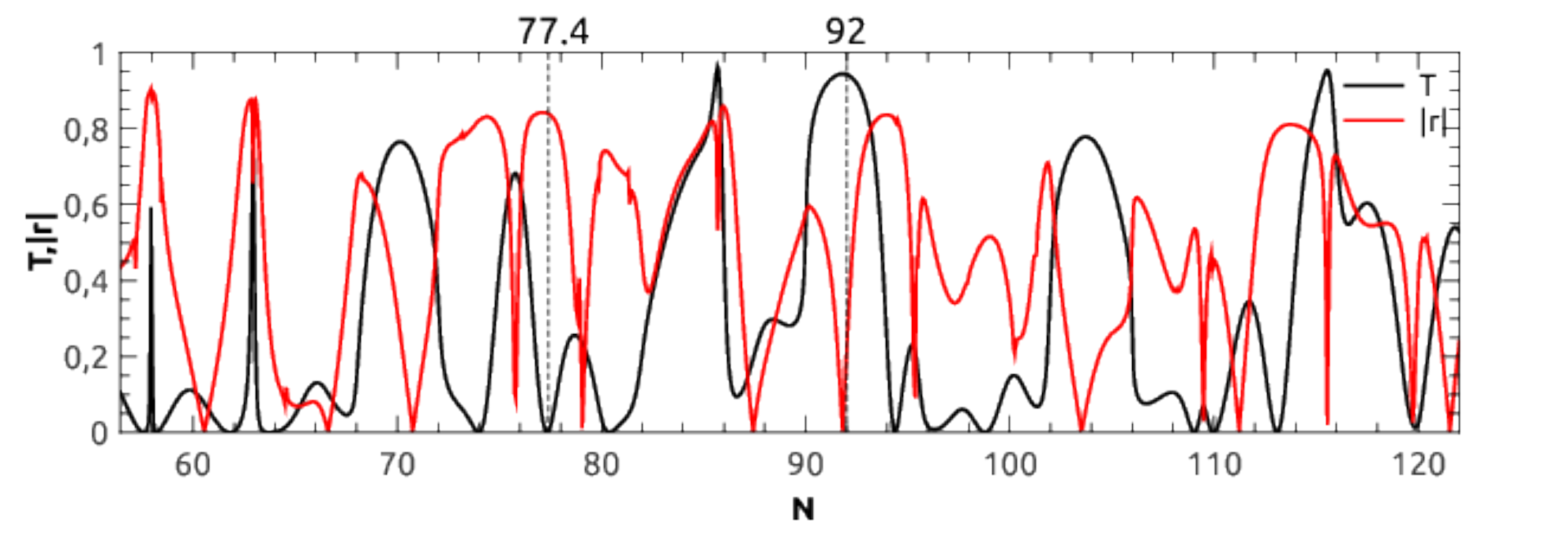}
          }
\caption{Same as Fig. \ref{todn} but for an asymmetric cavity of Fig. \ref{ns77}.}
\label{sns}
\end{figure}

\begin{figure}[ht!]
\hbox{
         	   \includegraphics[width=90mm]{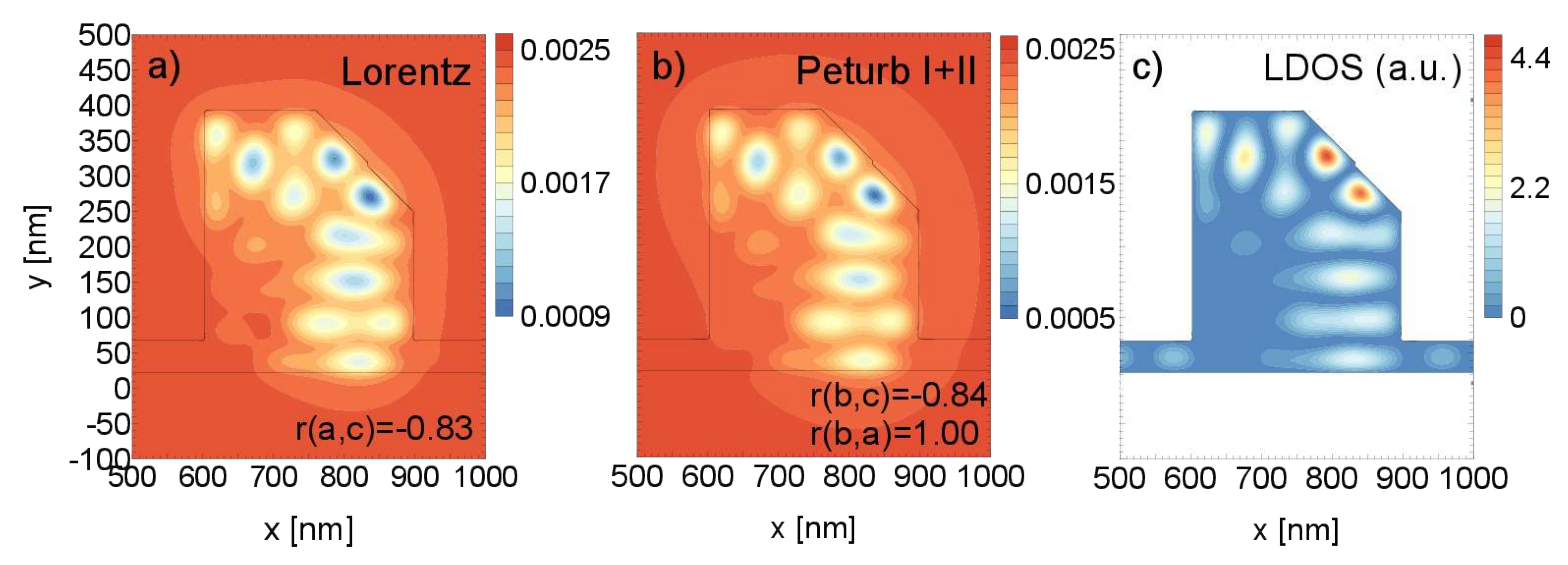}
          }
\caption{The results for $N=77$ electrons
and the asymmetric cavity obtained with a cut introduced to the right upper corner of the structure.
(a) The transfer probability map as obtained with the Lorentz tip potential, with the width of $d=4$ nm, and $U_0$
equal to 10\% of the kinetic Fermi energy. (b) The transfer probability obtained with the Lipmann-Schwinger equation up to the second correction.\cite{lpst}
(c) The local density of states. }
\label{ns77}
\end{figure}

\begin{figure}[ht!]
\hbox{
         	   \includegraphics[width=90mm]{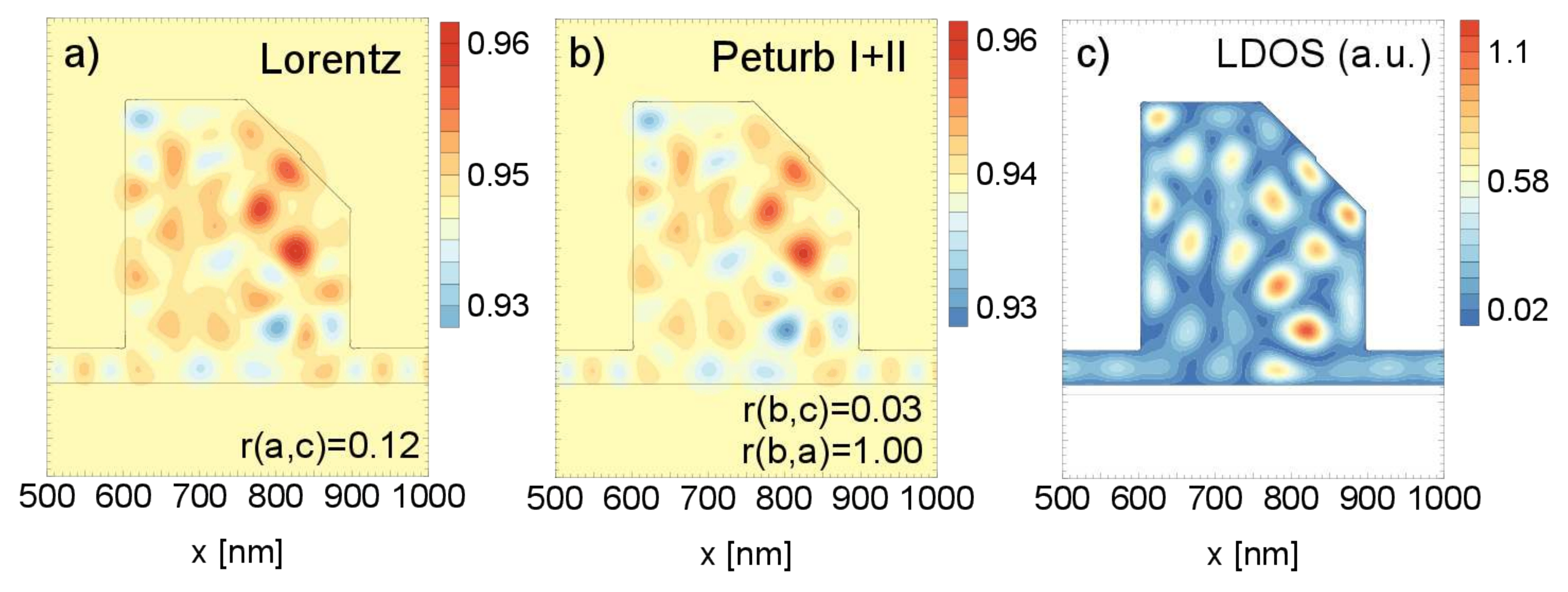}
          }
\caption{
Same as Fig. \ref{ns77} only for $N=92$ electrons.}
\label{ns92}
\end{figure}

The results presented so far were obtained for a symmetric structure. In the experiment the
asymmetry is inevitable and the present study requires generalization to a non-symmetric case.
The $T(N)$ and the correlation factor of LDOS to conductance maps obtained with the Lorentz
tip potential of width $d=4$ nm  and a height of $U_0=0.1 \frac{\hbar^2k^2}{2m_\mathrm{eff}}$ are displayed in Fig. \ref{sns} for the asymmetric cavity of Fig. \ref{ns77}.
We can see that, whenever  $T$ falls to zero -- as a result of the Fano resonance
with the states localized within the cavity -- one finds an increase of the correlation factor $r$ to about 0.8 or higher
[see the result of Fig. \ref{ns77}].
On the other hand for flat maxima of $T$ -- when the system is transparent for electrons -- or
in other words -- when there are no cavity-localized states for a given energy -- the correlation factor drops
to distinctly lower values. As a representative examples we plotted in Fig. \ref{ns77} and Fig. \ref{ns92} the conduction maps
for $N=77$ nad $N=92$ electrons for which a dip and a maximum of $T$ is obtained, respectively.
For the dip at $N=77$ a large correlation of $r=-0.83$ is found between the LDOS and the conductance map.
For the maximum at $N=92$ -- as seen above for symmetric cavity of Fig. \ref{ds}(e-f) --
the local extrema of $T$ and LDOS coincide but the maxima of one of the quantities
correspond to alternately a minimum or maximum of the other, hence the low value of $|r|=0.12$.

We checked that the conclusions reached in this work hold also when a small external magnetic field is applied perpendicular to the system.
For higher when the Zeeman splitting becomes
non-negligible as compared to the Fermi energy the Fano resonances for opposite spin orientations do not appear at the same
values of the electron densities, so indication of conditions for which the LDOS is
in a reliable manner imaged by conductance maps becomes difficult.

\section{Summary and Conclusions}

In summary, we have studied the correlation between the local density of states (LDOS) calculated
from the scattering wave functions of the
cavity at the Fermi level and the maps of the transfer probability $T$ in function of position of a
charged probe. We solved the Schroedinger-Poisson problem including the intrinsic screening of the
tip potential by the electron gas.
Our results indicate that the effective potential generated by the Coulomb charge
of the tip becomes short range when the tip approaches 2DEG at a close distance.

We studied both the conductance maps calculated with the Coulomb potential of the tip and
ansatz Lorentz potentials including the limit of point-like gentle perturbation introduced by the tip
for which the correlation of the LDOS to the conductance maps should be the closest.
We found as a general rule that the Fermi level wave functions can be quite precisely mapped by the model Lorentz tip potential of weak amplitude and short range
at the resonances involving states localized within the cavity. The resonances induce strong backscattering of the electron incoming from the Fermi level
and result in dips of the transfer probability falling to zero. Then, the correlation factor between LDOS and the electron transfer probability becomes
of the order of 0.8. Within the regions of flat plateaux of $T$, where no resonant localized states within the cavity are present, the conductance maps have no evident correspondence with the LDOS.

\acknowledgments
This work was supported by National Science Centre
according to decision DEC-2012/05/B/ST3/03290 and by
PL-Grid Infrastructure. Calculations were performed in ACK--
CYFRONET--AGH on the RackServer Zeus.

\end{document}